\documentclass[12pt,amsmath,amssymb]{article}
\pdfoutput=1
\usepackage{epsfig,amsmath,amssymb} 
\usepackage{graphicx}
\usepackage{dcolumn}
\usepackage{bm}
\usepackage{hyperref}
\usepackage{xcolor}
\hypersetup{
    colorlinks=true,       
    linkcolor=red,          
    citecolor=blue,        
    filecolor=magenta,      
    urlcolor=blue           
}
\usepackage{cite}
\usepackage[all]{hypcap}
\usepackage{cite}
\usepackage{epsfig}  
\usepackage{graphicx}               
\usepackage{url}
\usepackage{bm}
\usepackage{slashed} 
\usepackage{cancel}
\usepackage{braket}
\newcommand{\nc}{\newcommand}  
\newcommand{\mc}{\mathcal}
\nc{\beq}{\begin{equation}}  
\nc{\eeq}{\end{equation}}  
\nc{\beqa}{\begin{eqnarray}}  
\nc{\eeqa}{\end{eqnarray}}  
\nc{\bit}{\begin{itemize}}  
\nc{\eit}{\end{itemize}}

\title{
\vspace*{-2.3cm}
\begin{flushright}
\normalsize{
  }
\end{flushright}
\vspace{1.5cm}
\Large  
\textbf{Measuring the Black Hole Mass Spectrum from Redshifts of aLIGO Binary Merger Events } 
\vspace*{1.0cm}   
}

\author{Yang Bai, Vernon Barger and Sida Lu
\vspace{5mm}
\\
\normalsize\emph{Department of Physics, University of Wisconsin-Madison, Madison, WI 53706, USA}  
}
\date{}

\begin{document}  
\setcounter{page}{0}  
\maketitle  

\vspace*{1cm}  
\begin{abstract} 
The binary black hole merger events observed by the Advanced LIGO (aLIGO) and VIRGO collaboration can shed light on the origins of black holes. Many studies based on black hole stellar origins have shown a maximum mass for stellar black holes, which can be measured or constrained from the observed black hole mass distribution. In this paper, we point out that the redshift distribution of the observed merger events can provide complementary information for studying the black hole mass distribution, because the detectability correlates the event redshift to the black hole masses. With the improved sensitivity of aLIGO, a few dozen merger events may be obtained, for which we estimate that the maximum mass will be constrained to $10\,M_\odot$ accuracy.
\end{abstract} 
  
\thispagestyle{empty}  
\newpage  
  
\setcounter{page}{1}

\vspace{-2cm}

\section{Introduction}
\label{sec:intro}
In 2015, Advanced LIGO (aLIGO) observed the binary black hole (BBH) coalescence event of GW150914~\cite{Abbott:2016blz} in its first observation run. After that, there are four more BBH coalescence events discovered~\cite{Abbott:2016nmj,Abbott:2017vtc,Abbott:2017oio,Abbott:2017gyy} and one binary neutron star coalescence event~\cite{TheLIGOScientific:2017qsa}. More observations can be expected to be made after aLIGO finishes its final update, together with VIRGO~\cite{TheVirgo:2014hva}  and the joining of new  future detectors~\cite{Aso:2013eba,Unnikrishnan:2013qwa}. There is no doubt that we are now in a new era to study the BBH systems, and many questions about black holes can be investigated or examined from these abundant incoming data.

Among many topics in this field, one of the most important questions is to understand the origins of the black holes, which may manifest itself in the mass or redshift distributions of these binary systems. For instance, though it is sometimes argued that black holes with a mass of around $30\,M_\odot$ are larger than the expectation of supernovae explosions and stellar evolutions, it has been shown that these heavy black holes can still have a stellar origin~\cite{Belczynski:2016obo}. If this is the case, it would be natural to assume that the mass distribution of black holes follows the stellar initial mass function (IMF), which is a one-parameter power law for the black hole mass above the solar mass~\cite{Kroupa:2000iv}. Although additional stellar evolution processes could modify some simple correlation between the black hole and IMF mass functions. Anyhow, by fitting the mass distribution of the black holes in the observed BBH coalescence events, or by checking the consistency of other quantities after assuming the power-law IMF for the black holes, one could determine if the black holes observed by aLIGO and VIRGO are formed from the evolution of two stars or some other mechanisms~\cite{Nakamura:1997sm,deMink:2016vkw,Rodriguez:2016avt}.

Stars are usually divided into three different populations based on their metallicities. Metal-rich and metal-poor stars are usually referred to as Population I and II (Pop I/II) stars, while Populations III (Pop III) stars are known to be of virtually no metal. For black holes from Pop I/II star remnants, there would be a ``cap", or a maximum value, on their masses due to the mass loss of their progenitor stars by stellar winds. How strong these stellar winds are highly depends on the metallicity of the stars~\cite{Belczynski:2009xy}. Lower metallicity would in general lead to weaker wind and a heavier black hole, because low metal contents have a smaller opacity, enable easier radiation transportation, reduce the radiation momentum transfer and hence mass loss from surface. The heaviest stellar black hole observed so far, ignoring the remnants of BBH coalescence, is the primary black hole of GW150914 with a mass of $36^{+5}_{-4}\,M_\odot$. Using the lower value, $32\,M_\odot$, as a limitation, simulations have shown that the metallicity of its progenitor should be at most 1/2 of the solar metallicity~\cite{Spera:2015vkd} or even smaller than 1/4 of the solar metallicity~\cite{Belczynski:2009xy}. By fitting the BBH distributions with different mass-cap spectra to the observed events, one can set a better bound on the stellar black hole mass cap, and therefore obtain more information about the progenitors and formation environments of black holes.

Though Pop III stars have also been considered to be possible progenitors of the BBH coalescence events~\cite{Kinugawa:2014zha}, we do not consider this possibility here, because the merger rate density of Pop III binaries are significantly smaller than that of Pop I/II binaries~\cite{Belczynski:2016ieo}. However, if black holes from Pop III stars are detected, it would lead to a very interesting scenario,  in which the black hole mass spectrum is anticipated to have a mass gap rather than a simple mass cap. This is due to the so-called (pulsational) pair-instability supernovae [(P)PISNe]~\cite{Heger:2001cd}, for which the energetic photons can convert into a pair of electron and positron and change the radiation pressure. Numerical calculations have shown that the stellar black hole mass gap is $52-133\,M_\odot$~\cite{Woosley:2016hmi}. It will be interesting to see if the future events from aLIGO and VIRGO can establish the mass-cap or mass-gap spectrum. For the analysis in our paper, we will just take a phenomenological approach without being restricted to a specific type of stars. 

The mass distribution of the BBH system has been studied in a data-driven way by several recent papers from fitting physically motivated phenomenological parameters into observed results or pseudo events from a Monte Carlo simulation. For example, the studies in Ref.~\cite{Kovetz:2016kpi} used Fisher analysis to estimate the efficiency of constraining the model parameters with future sensitivity. Based on the mass distributions of the first three significant BBH coalescence events and the less significant event LVT151012~\cite{TheLIGOScientific:2016pea}, the authors in Ref.~\cite{Fishbach:2017zga} performed Bayesian parameter inference, and set constraints on the power-law index and the mass cap of the black holes. Similarly, the authors of Ref.~\cite{Talbot:2018cva} have tested a different mass spectrum with an accumulation of primary black hole masses at around $40\,M_\odot$ due to (P)PISNe, and performed Bayesian parameter inference to study the mass distribution. In this paper, we will use a similar phenomenological mass distribution model as in Ref.~\cite{Fishbach:2017zga} and perform a Kolmogorov-Smirnov (KS) test~\cite{smirnov1948} to constrain the maximum value of the stellar black hole masses based on the five observed events~\cite{Abbott:2016blz,Abbott:2016nmj,Abbott:2017vtc,Abbott:2017oio,Abbott:2017gyy}.

In addition to using the observed black hole mass distribution to learn the stellar black hole mass spectrum, we want to point out that one could also use the redshift distribution of the observed BBH coalescence events $\mc{R}_d(z)$ to constrain the mass spectrum parameters. This point has been ignored in previous studies and we will show that the observed redshift distribution can provide complimentary information and further constrain the mass cap or gap. The redshift distribution of observed events deserves more attention for several reasons. First, it is directly related to the redshift distribution of the binary merger rate density $R(z)$, and thus would reflect the information about the metallicity of the environment~\cite{Dominik:2013tma}. Moreover, for a fixed detector sensitivity, the maximum detectable redshift $z_{\rm max}$ of the detector relies on the masses of the two black holes in the BBH system. So, the redshift distribution of the observed events has a strong correlation with the black hole mass spectrum, and hence can be used to measure or constrain the mass-spectrum parameters.

For the first five observed events by aLIGO and VIRGO, four of them have the inferred redshift from the luminosity distance to be around $z \sim 0.1$. Though this peaked feature was later smeared out by the new events during aLIGO's second observation run\cite{LIGOScientific:2018mvr}. It is interesting to see how the detector detectability can work together with a certain mass spectrum to predict a peaked distribution function in $z$. For the upgraded aLIGO detector with reduced noise, the location of the peak in $z$ distribution will be shifted to a higher value when the binary system with heavier black holes is detected. On the other hand, this peak structure around $z\sim 0.1$ could have its origin as an intrinsic property of the merger rate density $R(z)$ from the formation history of the BBH systems. Future BBH merger events from aLIGO and VIRGO will provide a conclusive answer to the above two explanations. 

This paper is organized as follows. In Section~\ref{sec:SNR}  we summarize the information of all five observed BBH merger events, the detectability of a merger event on a single detector, and provide a parametrized expression for the background noise, signal and maximum detectable redshift as a function of black hole masses. In Section~\ref{sec:anticipated}, we calculate the predicted distributions in redshift and the primary black hole mass for different mass spectra. In Section~\ref{sec:five-events}, we use the KS test to constrain the mass-spectrum parameters based on the five observed events, while in Section~\ref{sec:future} we briefly investigate the future sensitivity of the upgraded aLIGO detectors on constraining the mass-spectrum parameters. In Section~\ref{sec:ten-events} we update our analysis to take into account the new observed and recognized events. Finally, we conclude in Section \ref{sec:conclusion}.

\section{Detectability based on aLIGO}
\label{sec:SNR}
For the five BBH merger events with large observational significance, the black hole masses and observed redshifts are listed in Table~\ref{tab:data}.  
\begin{table}[t!]
\begin{center}
\renewcommand{\arraystretch}{1.5}
\begin{tabular}{cccccc}
\hline\hline
Event & GW150914 & GW151226 & GW170104 & GW170608 & GW170814\\
$m_1/M_{\odot}$ & $35.6^{+4.8}_{-3.0}$ & $13.7^{+8.8}_{-3.2}$ & $31.0^{+7.2}_{-5.6}$ & $10.9^{+5.3}_{-1.7}$ & $30.7^{+5.7}_{-3.0}$\\
$m_2/M_{\odot}$ & $30.6^{+3.0}_{-4.4}$ & $7.7^{+2.2}_{-2.6}$ & $20.1^{+4.9}_{-4.5}$ & $7.6^{+1.3}_{-2.1}$ & $25.3^{+2.9}_{-4.1}$\\
$z$ & $0.09^{+0.03}_{-0.03}$ & $0.09^{+0.04}_{-0.04}$ & $0.19^{+0.07}_{-0.08}$ & $0.07^{+0.02}_{-0.02}$ & $0.12^{+0.03}_{-0.04}$\\
$E_{\rm rad}/(m_1 + m_2)$ & 0.046 & 0.047 & 0.043 & 0.049 & 0.048\\
\hline
Event & GW151012 & GW170729 & GW170809 & GW170818 & GW170823\\
$m_1/M_{\odot}$ & $23.3^{+14.0}_{-5.5}$ & $50.6^{+16.6}_{-10.2}$ & $35.2^{+8.3}_{-6.0}$ &  $35.5^{+7.5}_{-4.7}$ & $39.6^{+10.0}_{-6.6}$\\
$m_2/M_{\odot}$ & $13.6^{+4.1}_{-4.8}$ & $34.3^{+9.1}_{-10.1}$ & $23.8^{+5.2}_{-5.1}$ & $26.8^{+4.3}_{-5.2}$ & $29.4^{+6.3}_{-7.1}$\\
$z$ & $0.21^{+0.09}_{-0.09}$ & $0.48^{+0.19}_{-0.20}$ & $0.20^{+0.05}_{-0.07}$ & $0.2^{+0.07}_{-0.07}$ & $0.34^{+0.13}_{-0.14}$\\
$E_{\rm rad}/(m_1 + m_2)$ & 0.041 & 0.057 & 0.046 & 0.043 & 0.048\\
\hline\hline
\label{tab:data}
\end{tabular}
\caption{A summary of the properties of the ten observed BBH merger events, with the data taken from~\cite{LIGOScientific:2018mvr}. The five events at the bottom of the table are the recently reported/recognized ones. The last row is the fraction of radiated energy in the total mass.}
\end{center}
\end{table}
It is obvious that the maximum mass for the stellar black holes $M_{\rm max}$ should be at least above the largest observed black hole mass, and hence a lower bound can be set: $M_{\rm max} \gtrsim 36\,M_\odot$. In this paper, we are interested in the upper bound on $M_{\rm max}$ based some reasonable black hole mass spectrum and the detection sensitivity. We want to show that the observed redshift for the five events could have a strong correlation with the upper value of $M_{\rm max}$.
%
%

Following Ref.~\cite{Flanagan:1997sx}, we use the signal-to-noise ratio (SNR) to determine the detectability of one BBH merger event with $m_1$, $m_2$, $z$ and other parameters for the detectors at the Hanford and Livingston. For the three phases in time-domain: inspiral, merger, and ringdown, one can separate them by two characteristic frequencies. The separating frequency value between the inspiral and merger phases is
\begin{align}
f_{\rm merge} = \epsilon_i/M\approx 136\,\mbox{Hz}\times[60M_\odot/M]  ~~,
\end{align}
for $\epsilon_i=0.04$ in units of $G_N=1$ and $c=1$. Here, the total mass $M\equiv m_1 + m_2$.  The separating frequency between the merger and ringdown phases has $f_{\rm ringdown} = [1-0.63(1-a_f)^{3/10}]/(2\pi M) \approx 295\,\mbox{Hz}\times(60M_\odot/M)$ for the dimensionless spin $a_f=0.70$, which is the weighted average of the five merger events.  

\begin{figure}[thb!]
\begin{center}
\includegraphics[width=0.6\textwidth]{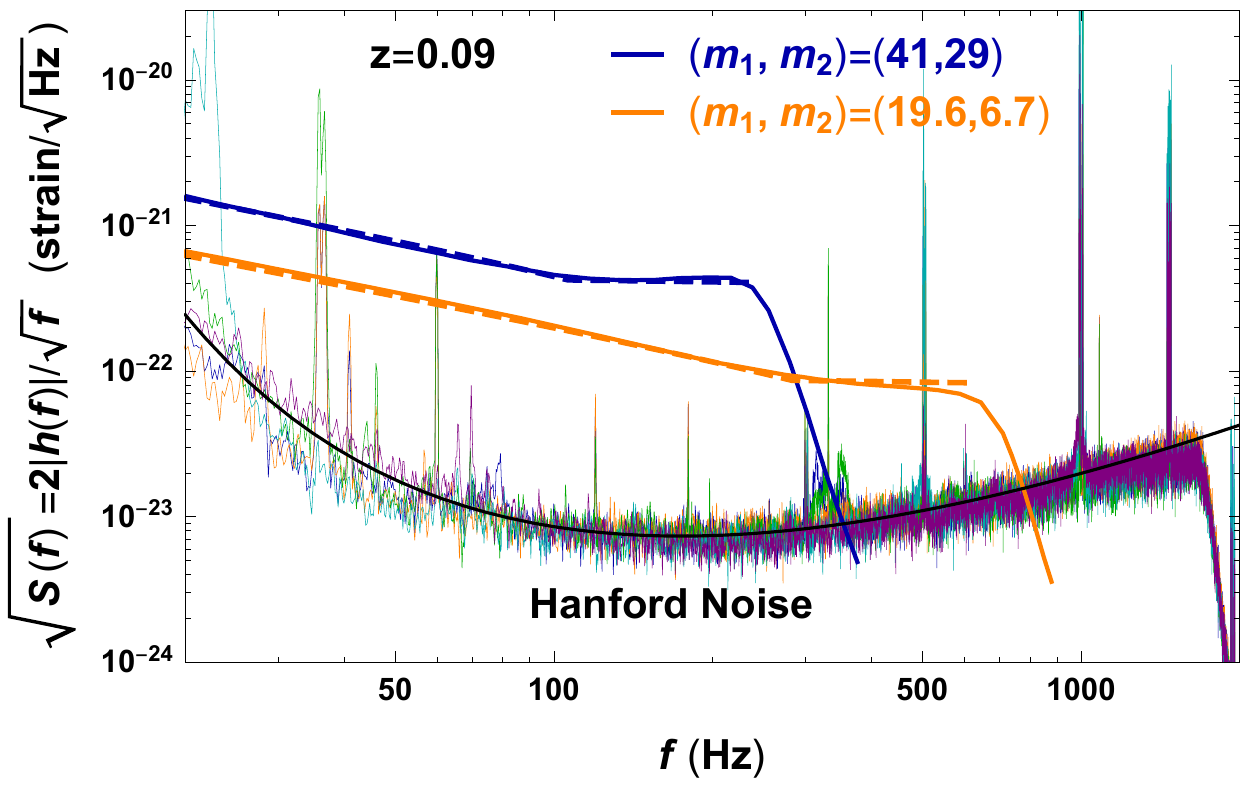}
\vspace*{-0.3cm}
\caption{The strain per $\sqrt{f}$ as a function of gravitational wave frequency. The upper solid lines are the signal templates from the aLIGO collaboration (from the Gravitational Wave Open Science Center)~\cite{ligo:tutor}, while the dashed lines are based on the approximate functions in \eqref{eq:inspiral-merge} with $z=0.09$. All black hole masses are in $M_\odot$. The spiky and thin lines are the background noise at the Hanford detector during the five observed events, while the black and solid line is from the fitted function in \eqref{eq:noise-fit}. }
\label{fig:signal-fit}
\end{center}
\end{figure}

Rather than directly use the approximate formulas in Ref.~\cite{Flanagan:1997sx}, we use the signal templates from the aLIGO collaboration to determine a simple empirical form of the spectral energy density~\cite{ligo:tutor}. The spectral energy density for the inspiral and merger phases in terms of the GW frequency at the source is given by
\beqa
\left(\frac{dE}{df_s}\right)^{\rm inspiral} &=& \frac{\pi^{2/3}}{3\,f_s^{1/3}}\,\mu\,M^{2/3} \left\{ 1 - \left(\frac{3}{2}+ \frac{\nu}{6}\right)(\pi\,f_s\,M)^{2/3}  \right\}\,,  \qquad \mbox{for}\quad f_s < f_{\rm merge} \,, \nonumber \\ 
\left(\frac{dE}{df_s}\right)^{\rm merge} &=& \left.\left(\frac{dE}{df_s}\right)^{\rm inspiral}\right|_{f_s=f_{\rm merge}} \times \left( f_s/f_{\rm merge}\right)^{0.9} \,, \qquad \mbox{for}\quad  f_{\rm merge}   \leq f_s < f_{\rm ringdown} \,,
\label{eq:inspiral-merge}
\eeqa
where the reduced mass is $\mu \equiv m_1 m_2 / M$ and the dimensionless quantity $\nu \equiv \mu/M$. 
For the inspiral phase, we keep the next-leading term in the post-Newtonian expansion~\cite{Will:1996zj}. Requiring this function to be continuous in frequency, one can fix the overall normalization during the merger phase. In Figure~\ref{fig:signal-fit}, we show a comparison of our fitted function and the templates used by the LIGO collaboration~\cite{ligo:tutor}, which shows a good agreement. For our semi-analytic approach, we will use the spectral energy density in Eq.~\eqref{eq:inspiral-merge} without spin effects. 

The ringdown phase has a small contribution to SNR with its frequency spectral form as
\beqa
\left(\frac{dE}{df_s}\right)^{\rm ringdown} =  \frac{\mathcal{A}^2_m\,f_{\rm ringdown}^2\,f_s^2\,M^2}{32\pi\,Q^2} \sum_{\pm}\left[(f_s \pm f_{\rm ringdown})^2 + \frac{f_{\rm ringdown}^2}{4\,Q^2}\right]^{-2} \,,
\eeqa
where the dimensionless parameter $Q=2(1-a_f)^{-9/20}\approx 3.4$ for $a_f=0.7$ and the dimensionless amplitude parameter $\mathcal{A}_m=\mathcal{A} \times 16\,\nu^2$. In the limit of $Q \gg 1$, the function during the ringdown phase can be well approximated by a delta function: $\frac{1}{8}\mathcal{A}^2_mQM^2f_{\rm ringdown}\delta(f_s - f_{\rm ringdown})$~\cite{Flanagan:1997sx}. 

The characteristic gravitation-wave amplitude at a local detector for a source at a redshift $z$ is given by \cite{Flanagan:1997sx}
\beqa
h_s(f)^2 \equiv \frac{2(1+z)^2}{\pi^2 D_L(z)^2} \, \frac{dE}{df}[(1+z) f] \,,
\eeqa
as a function of frequency, with $D_L(z)$ as the luminosity distance of the source. It has the form of $D_L(z) = (1+z)D_C(z)$ with the co-moving distance as $D_C(z) = c/H_0 \times \int^z_0 [\Omega_{m}(1+z')^3 + 1 - \Omega_{m}]^{-1/2} dz'$ during the matter-dominated universe. We will use $H_0=67.8~\mbox{km/s/Mpc}$~\cite{Ade:2015xua} (a higher value of $H_0$ from local measurement~\cite{2018arXiv180101120R} can lead to a simple rescale of SNR) and $\Omega_{m}=0.308$~\cite{Ade:2015xua}. Here, the relation between the frequency at the source $f_s$ and the detected $f$ is $f_s = (1+z) f$.  For a small value of $z$, the strain per $\sqrt{f}$ for the signal has an approximately simple power-law behavior
\beqa
\sqrt{S_s(f)} = 2|h_s(f)|/\sqrt{f} \approx \left\{
\begin{array}{ll}
 \dfrac{2\sqrt{{2\over3}} \, H_0\,\mu^{1/2}\,M^{1/3} }{\pi^{2/3}\,f^{2/3}\, z}  & \mbox{for}\quad f< f_{\rm merge}  \,, \vspace{2mm} \\
 \dfrac{5.5\, H_0\,\mu^{1/2}\,M^{0.95} }{f^{0.05}\, z}  & \mbox{for}\quad  f_{\rm merge}   \leq f < f_{\rm ringdown} \,.
\end{array}
\right.
\eeqa
So, during the merging phase and because of the smaller power in $f$, the quantity $\sqrt{S_s(f)}$ is approximately flat in $f$, as can be seen in Figure~\ref{fig:signal-fit}.

Integrating out the frequency, we have the ratio of the radiated energy over the total black hole mass as
\beqa
\dfrac{E_{\rm rad}}{M} &=& r_{\rm inspiral} + r_{\rm merge} + r_{\rm ringdown} \approx 0.1\,\nu \,+\, 0.1\,\nu\, +  \, 0.038\,\mathcal{A}^2 \,,
\label{eq:noise-fit}
\eeqa
for $a_f = 0.7$. As estimated in Ref.~\cite{Flanagan:1997sx}, the maximum value of $\mathcal{A}$ is taken to be $\mathcal{A}=0.4$. For a smaller value of $\mathcal{A}$, the contribution to the radiation energy during the final ringdown phase is negligible. Interestingly, using our parametrization in Eq.~\eqref{eq:inspiral-merge}, the radiated energies during the inspiral and merger phases are approximately equal. For the mass ratios of the observed five events in Table~\ref{tab:data}, the parameter $\nu$ ranges from 0.22 to its maximum value 0.25.  So, the radiated energy fraction is from 0.045 to 0.050 which is in good agreement with the reported values by the LIGO collaboration (see Table~\ref{tab:data}). 
 
For the background noise, we take the less sensitive detector at Hanford and require some minimum value of SNR around 8 to claim a detection of one merger event. For the few observed events, they have similar noise strains, which can be well parametrized and fitted by 
\beqa
\sqrt{S_n(f)} &=& 2|h_n(f)|/\sqrt{f}   \approx 6.7\times 10^{-34} \, f^{2.24}\,e^{59.6/\ln{f}}\,,
\label{eq:noise-fit}
\eeqa
with $f$ in Hz and $S(f)$ in strain per $\sqrt{\mbox{Hz}}$. In Figure~\ref{fig:signal-fit}, we show a comparison of the Hanford background noise and our fitted function. Comparing the signal and the background noise, one can see that as the summed black hole mass increases, the background noise for the corresponding frequency range increases faster. So, the detector sensitive should be peaked at around $200\,M_\odot$ for the summed mass.

The signal-to-noise ratio squared for a randomly oriented source has
\beqa
\left(\frac{S}{N}\right)^2 = \frac{2(1+z)^2}{5\,\pi^2 D_L(z)^2}\int^\infty_0 df \,\frac{1}{f^2\,S_n(f)} \frac{dE}{df}[(1+z) f]  \,,
\eeqa
where the extra factor of 5 comes from the root-mean-square average of signal amplitudes over different possible orientations of the source and interferometer~\cite{Finn:1992xs}. In using the above equation to estimate the sensitivity at the aLIGO detectors, we choose the frequency integration range from 20 Hz to 2000 Hz~\cite{Abbott:2017vtc}. To claim an observation at one specific detector, we also require a minimum value of $\mbox{SNR}_{\rm min}$ with $S/N > \mbox{SNR}_{\rm min} =8$~\cite{TheLIGOScientific:2016zmo}.

\begin{figure}[thb!]
\begin{center}
\includegraphics[trim={0 0.5cm 0 0}, clip, width=0.6\textwidth]{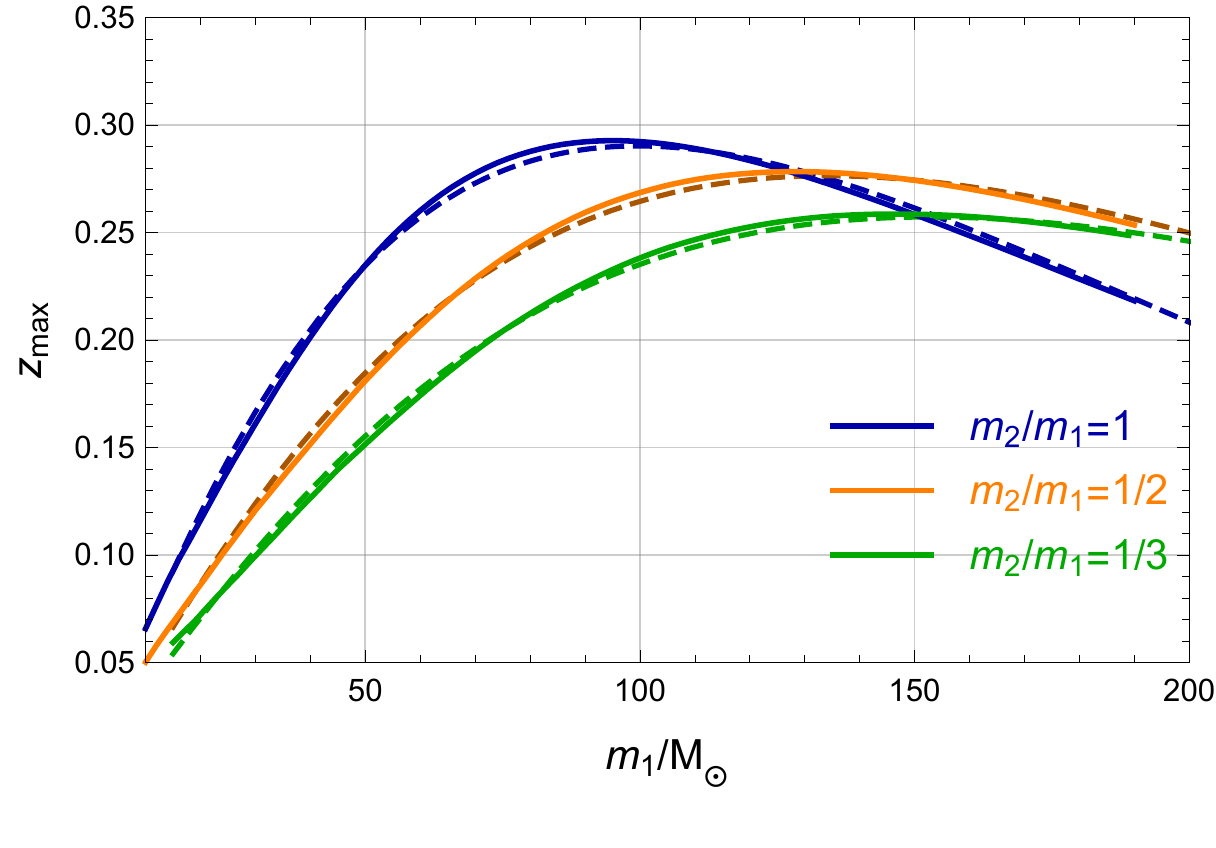}
\caption{The maximum reach of redshift, $z_{\rm max}$, for different $m_1$ and $m_2$ and based on the Hanford background noise.}
\label{fig:zmax-fit}
\end{center}
\end{figure}

For given values of $m_1$ and $m_2$ and requiring detectability at the Hanford detector, a maximum value of redshift, $z_{\rm max}(m_1, m_2)$, can be calculated numerically. Again, we can use an empirical function to fit the numerical answers. For a wide range of $m_1$ and $m_2$, we have
\beqa
z_{\rm max}(m_1, m_2) &\approx& \exp\left\{-4.83 + 0.642\,\ln{\left[\frac{(m_1\,m_2)^{0.9}}{(m_1 + m_2)^{0.3}}\right] } \right.\nonumber \\
&& \hspace{0.8cm} \left.-10^{-6}\times\left[4.39+1.11\left(\frac{m_1}{m_2}+\frac{m_2}{m_1}\right)^{0.91}\right]\,\ln^{6.1}{\left[\frac{(m_1\,m_2)^{0.9}}{(m_1 + m_2)^{0.3}}\right] } \right\} \,.
\eeqa
Here, both black hole masses, $m_1$ and $m_2$, are in the unit of the solar mass $M_\odot$. 
In Figure~\ref{fig:zmax-fit}, we show a comparison of $z_{\rm max}$ as a function of $m_1$ for different ratios of $m_2/m_1$ with the fitted function. 
From the blue line of Figure~\ref{fig:zmax-fit}, one can see that the largest $z_{\rm max}$ happens when $m_1=m_2\approx 85\,M_\odot$ with a value of around 0.29. This is simply due to the fact that the background noise has a minimum value around 200 Hz (see Figure~\ref{fig:signal-fit}). 

\section{Anticipations from different black-hole mass spectra}
\label{sec:anticipated}
The main purpose of our paper is to learn some black hole  properties from the observed BBH merger events. The observed black hole masses can easily tell us the range and spectrum of black hole masses, as studied in Refs.~\cite{Fishbach:2017zga,Kovetz:2016kpi}. In this paper, we also point out that the observed distribution in $z$ can aid in determining the black hole mass function. Concentrating on the three BBH merger property parameters, $m_1$, $m_2$ and $z$,  the general detection rate of BBH merger events including the detectability effect can be expressed as
\beqa
\frac{d^3 \mathcal{R}_d(m_1, m_2, z)}{d m_1\, d m_2\,d z} 
&=& p(m_1, m_2)\,R(z)\, \frac{d V_c}{dz} \frac{dt_s}{dt} \, \Theta[z_{\rm max}(m_1, m_2) - z]~~.
\label{eq:triple-diff}
\eeqa
Here, $\Theta(x)$ is Heaviside function to encode the detectability; $d V_c/dz\propto D_{C}^2(z)/H(z)$ is the comoving volume at a given redshift with $H(z)$ the redshift-dependent Hubble parameter; $dt_s/dt_0 = 1/(1+z)$ is the time dilation factor between the source and the detector. In general, the black hole merging events may depend on a non-factorizable function, $f(m_1, m_2, z)$, of masses and redshift. For instance, the location of a cap or gap of black hole masses may depend on metallicity and redshift of the protostars~\cite{Woosley:2002zz}. In our later analysis, we make a simplified assumption with a factorizable $f(m_1, m_2, z)=p(m_1, m_2)\,R(z)$. 

The binary merger rate density $R(z)$ is a function of the redshift of the BBH system. For the stellar origin of the black holes, we parametrize the simulation results of metallicity around $0.1\,Z_\odot$ binaries in Ref.~\cite{Dominik:2013tma} as 
\beqa
R(z)= R_0\,e^{\gamma_z\,(1+z)} \,,
\eeqa
with a fitted result $\gamma_z \approx 1.1$ in our later analysis. As one can see from Figure~\ref{fig:zmax-fit}, the current aLIGO has sensitivity up to $z\approx 0.29$, which has $R(z)$ insensitive to the errors of $\gamma_z$. 

For the mass distribution function, $p(m_1, m_2)$, we will consider the following general form to describe a power-law plus a ``gap" shape
\beqa
p(m_1, m_2) &\propto& m_1^{-\alpha} \,m_2^\beta\, \qquad \mbox{for}\,\quad M_{\rm min} \leq m_i <  M^{\rm low}_{\rm gap} 
\quad \mbox{or} \,\quad M^{\rm high}_{\rm gap} \leq m_i \,. 
\label{eq:mass-parametrization}
\eeqa
In the above parametrization, $m_1(m_2)$ is the heavier(lighter) black hole mass of the BBH system. For the stellar black hole masses, there is a lower stellar black hole mass limit, $M_{\rm min}$, which will be taken as $M_{\rm min}=5\,M_\odot$~\cite{Ozel:2010su} in the later analysis of this paper. The power $\beta$ indicates the sensitivity on the lighter black hole mass. When $\beta=0$, used in the aLIGO collaboration analysis~\cite{TheLIGOScientific:2016pea}, the black-hole binary system mainly depends on $m_1$. 

The heavier black hole masses could follow its dependence from the IMF because of their stellar origin. Using the averaged power law index for the galactic-field initial masses in Ref.~\cite{Kroupa:2000iv}, the power is 
\beqa
\alpha= 2.3 \pm 0.7 ~~,
\eeqa
within the 99\% confidence interval. In our later analysis, we will focus on this range of power law index and show how our conclusions depend on this power. 

The black hole masses are also anticipated to have a mass ``cap" or a mass ``gap". In Eq.~\eqref{eq:mass-parametrization}, we have introduced two parameters, $M^{\rm low}_{\rm gap}$ and $M^{\rm high}_{\rm gap}$, to denote the range of allowed black hole masses: $m_2\leq m_1 \leq M^{\rm low}_{\rm gap}$ or  $M^{\rm high}_{\rm gap}\leq m_2 \leq m_1$ or $m_2 \leq M^{\rm low}_{\rm gap}<  M^{\rm high}_{\rm gap}\leq m_1$. For an extremely large value of the higher-end of the mass gap, $M^{\rm high}_{\rm gap}\rightarrow \infty$, the mass spectrum becomes simply a power-law with a mass cap and one can identify 
\beqa
M_{\rm max} \equiv M^{\rm low}_{\rm gap} ~~.
\eeqa
%
Depending on the underlying models for black hole formation and metallicity of the environment, this mass cap could range from $\sim 30\,M_\odot$ to $\sim 100\,M_\odot$~\cite{Belczynski:2009xy}. Given the primary mass $m_1 = 36.2^{+5.2}_{-3.8}$ for the event GW150914, the parameter $M_{\rm max}$ is bounded from below, {\it e.g.,} $M_{\rm max} \geq 36\,M_\odot$. However, the upper limit on $M_{\rm max}$ is not obvious and may rely on some statistical inferences. The existing studies in Refs.~\cite{Kovetz:2016kpi,Fishbach:2017zga} have proposed to use the observed primary masses to obtain an upper bound on $M_{\rm max}$. In this paper, as well as performing a similar analysis, we show that the redshift distribution of the observed events can provide complimentary information to the study of the mass spectrum. 

Furthermore, the possible existance of a mass gap for the range of around $50\,M_\odot - 130\,M_\odot$ has been suggested in Ref.~\cite{Woosley:2002zz,Woosley:2016hmi,Gilmer:2017zef,Spera:2017fyx} based on the (P)PISNe. In this paper, we show how different values of $M^{\rm high}_{\rm gap}$ change the observed $m_1$ and $z$ distributions. When additional events are accumulated by aLIGO, the mass spectrum parameters, $M_{\rm max}$ or $M^{\rm high}_{\rm gap}$, can be measured and thereby provide information about the underlying dynamics of stellar black hole formation.

\begin{figure}[htb!]
\begin{center}
\includegraphics[width=0.6\textwidth]{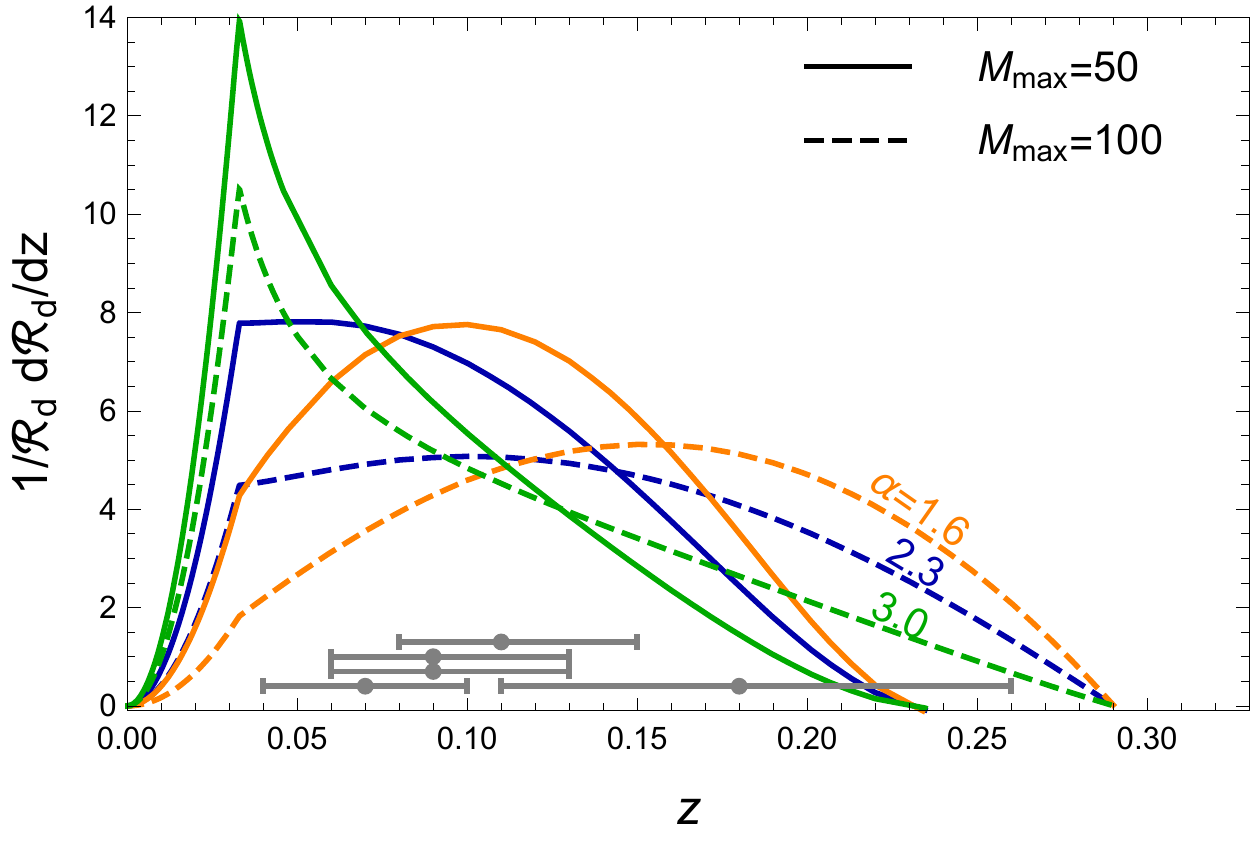}
\vspace*{-0.3cm}
\caption{The expected event distributions in redshift after taking into account detector sensitivity. The solid(dashed) lines are for $M_{\rm max}=50(100)\,M_\odot$, respectively. Here, $M_{\rm min}=5\,M_\odot$ and $\beta=0$. The redshifts with errors of the five observed events are shown at the bottom in gray horizontal lines.}
\label{fig:z-distribution}
\end{center}
\end{figure}

Based on the current Hanford sensitivity, Figure~\ref{fig:z-distribution} shows the normalized redshift distribution based on a power-law mass spectrum with a mass cap with $\beta=0$, $M_{\rm min}=5\,M_\odot$ for different values of $\alpha$ and $M_{\rm max}$. More specifically, we integrate out the two mass variables $m_1$ and $m_2$ by taking the detectability into account [see Eq.~\eqref{eq:triple-diff}]. Comparing the curves with $M_{\rm max}=50\,M_\odot$ and $100\,M_\odot$, one can see that a larger value of $M_{\rm max}$ has a wider range of $z$ distribution and has a larger averaged $z$. This behavior can be easily understood from the maximum reach of redshift in Figure~\ref{fig:zmax-fit} for different values of $m_1$. Once $M_{\rm max}$ is above $85\,M_{\odot}$, the observed events can have a nonzero probability of reaching the largest value of about 0.29 for $z_{\rm max}$. For an even larger value of $M_{\rm max}$ above $100\,M_\odot$, the end point of $z$-distribution will not change, but the distributions will be shifted to slightly higher values of $z$. For $M_{\rm max}=50\,M_\odot$, the maximum value for the redshift of the observed events is around 0.23. So, if aLIGO observed an event with a precise redshift above 0.23, the parameter $M_{\rm max}$ has to be above $50\,M_\odot$. 

For a smaller value of $\alpha$, the mass spectrum is shallower, which implies more heavy black holes. As a consequence, more events with larger values of $z$ are anticipated. For a larger value of $\alpha$ and a small value of $M_{\rm max}$ (see the green and solid line Figure~\ref{fig:z-distribution} for instance), the $z$-distribution should be peaked at lower values, below around 0.05, which obviously does not match the five observed events. Therefore, we anticipate a lower bound on $M_{\rm max}$ for a large value of $\alpha$. 

We also note that there is a ``kink" structure at $z\approx 0.03$ in Figure~\ref{fig:z-distribution}, which is due to the step-function requirement of $m_{1,2} \geq M_{\rm min}=5\,M_\odot$. In principle, with enough events, one could also use the $z$-distribution to ``measure" the important spectrum parameters: $M_{\rm max}$ and $\alpha$.

\begin{figure}[htb!]
\begin{center}
\includegraphics[width=0.6\textwidth]{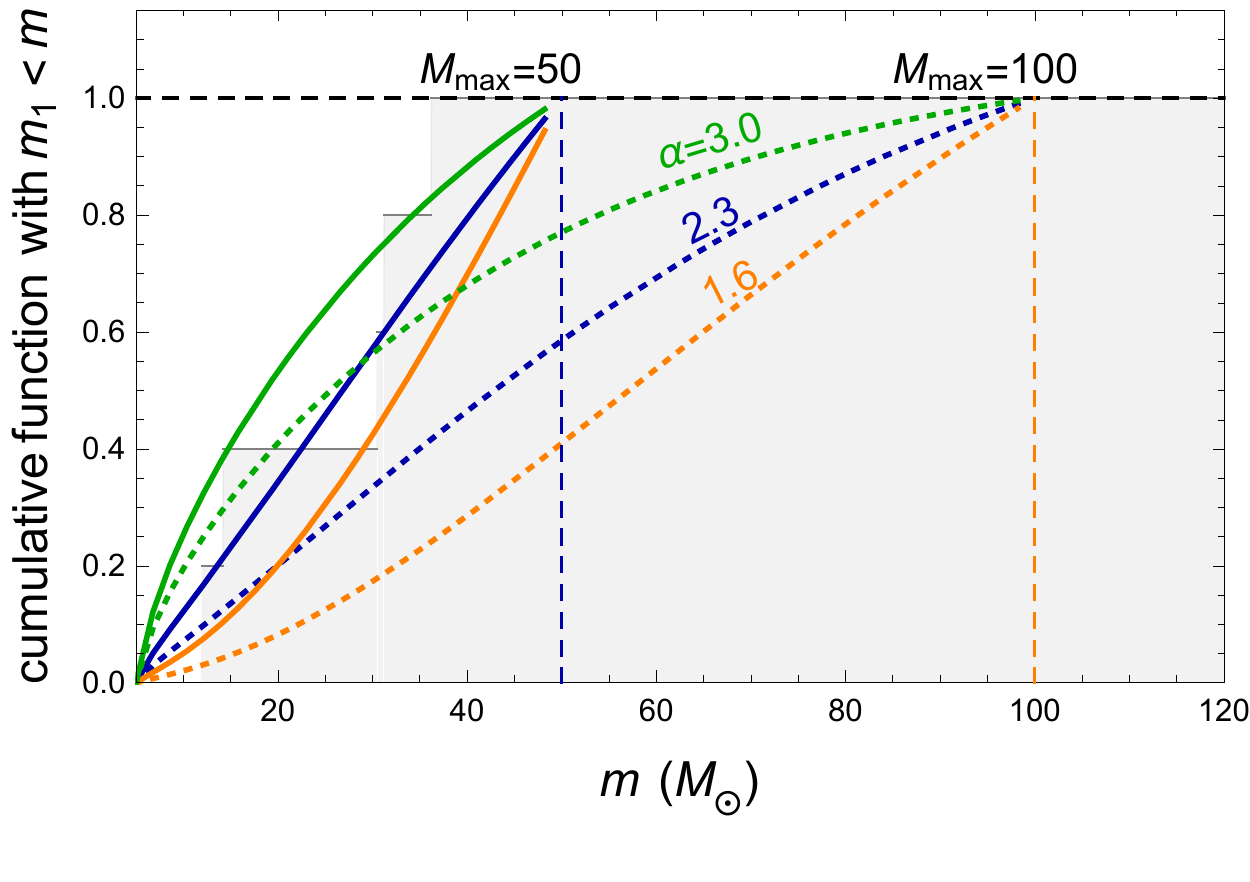}
\vspace*{-0.3cm}
\caption{Cumulative distribution function $\mathcal{R}_d^{-1}\int^m_{M_{\rm min}} dm_1 d\mathcal{R}_d/dm_1$ in the primary black hole mass for $M_{\rm min}=5\,M_\odot$, $\beta=0$ and different $\alpha$, $M_{\rm max}$. The corresponding distribution for the five observed events is shown in the gray and stair-like distribution.}
\label{fig:m1-distribution}
\end{center}
\end{figure}

Similarly, one could also integrate out the variables $m_2$ and $z$ using Eq.~\eqref{eq:triple-diff} to obtain the observed event distribution in $m_1$. For different values of $M_{\rm max}$ and $\alpha$, we show the cumulative probability distributions for different values of $m$ in Figure~\ref{fig:m1-distribution}. The corresponding cumulative distribution for the five observed events is shown in the gray stair-like histogram. For $M_{\rm max}=50\,M_\odot$, the model-predicted distributions are more or less agree with the observed distributions. For a shallower mass spectrum, one anticipates more events with heavier black hole masses. Absence of heavier black holes can therefore set an upper limit on $M_{\rm max}$. For instance, for $\alpha=1.6$ and $M_{\rm max}=50\,M_\odot$, we anticipate 70\% of events with a mass above $40\,M_\odot$, which is in contradiction to the current observed distribution. So, an upper bound on $M_{\rm max}$ should exist between $50\,M_\odot$ and $100\,M_\odot$.

\section{Constraints from the five observed  events}
\label{sec:five-events}
To quantify the possible limits on the mass-spectrum parameters, we use the simple KS test~\cite{smirnov1948} to quantify the difference between the model predicted and observed distributions. The test of statistics (TS) is defined as 
\beqa
\mbox{TS} \equiv \mbox{sup}_z \left| F_n^{\rm data}(z)   - F^{\rm model}(z)\right|\,,
\eeqa
where ``$\mbox{sup}$" means supremum or the maximum value of difference for any $z$. Here, $F_n^{\rm data}(z)$ as the cumulative distribution function for some data, while $F^{\rm model}(z)$ is the cumulative distribution function for models with some mass spectrum. In the case with a large number of observed events and if the model provides the right distribution, one should anticipate $\mbox{TS}\rightarrow 0$ for $n\rightarrow \infty$. To calculate the $p$-value for the consistency of the model with the observed data, one can randomly generate samples of pseudo-data based on the model distribution and then construct a TS distribution to calculate the probability of obtaining $\mbox{TS}^{\rm obs.}$.

\begin{figure}[htb!]
\begin{center}
\includegraphics[width=0.6\textwidth]{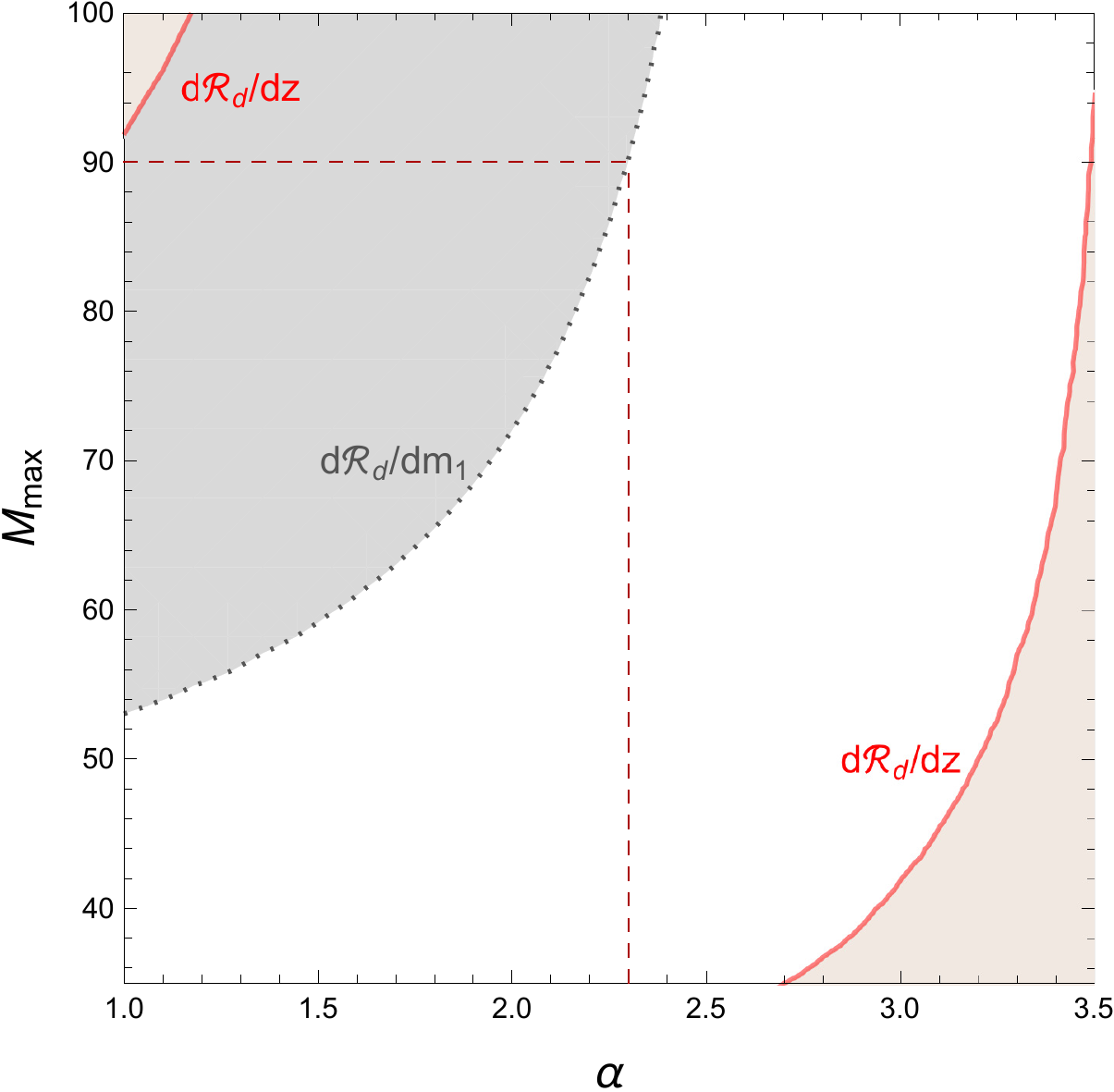}
\vspace*{-0.3cm}
\caption{The constraints on the power-law mass-spectrum parameters: $\alpha$ and $M_{\rm max}$, at 95\% confidence level (or the boundaries with 5\% $p$-value for the KS test) from the $d\mathcal{R}_d/dz$ distribution (light brown region) and $d\mathcal{R}_d/dm_1$ distribution (light gray region) of the observed five events.  Here, $M_{\rm min}=5\,M_\odot$ and $\beta=0$. 
}
\label{fig:p-value}
\end{center}
\end{figure}

Using the observed redshift's of the five observed events in Table~\ref{tab:data}, we show the region with the KS-test $p$-value less than 5\% in the light-brown region of Figure~\ref{fig:p-value} for different values of $M_{\rm max}$ and $\alpha$. Requiring the $p$-value above 5\%, or the 95\% confidence level, the current five events already constrain the mass-spectrum parameters. For the range of $2.4 \lesssim \alpha \lesssim 3.0$, the model parameter $M_{\rm max}$ is constrained to be above $36\,M_\odot$ to $42\,M_\odot$. For the central value of $\alpha=2.3$, there is no stringent constraint on $M_{\rm max}$ from the current five-event sample, just based on the $d\mathcal{R}_d/dz$ distribution. For a smaller value of $\alpha \sim 1.6$, there is a weaker upper bound, around $100\,M_{\odot}$, on $M_{\rm max}$. It is interesting to see that even with only five events, the observed $z$-distribution can already provide useful information for us to understand the black hole mass spectrum. 

For the observed $d\mathcal{R}_d/d{m_1}$ distribution, we show the 5\% $p$-value boundary in the gray region of Figure~\ref{fig:p-value}. For the central value $\alpha=2.3$, the $m_1$ distribution provides a more stringent constraint than that from the $z$-distribution. An upper limit on $M_{\rm max}$ is obtained to be around $90\,M_\odot$. For a shallower mass spectrum with $\alpha=1.6$, one has a more stringent constraint on $M_{\rm max}$ to be below $56\,M_\odot$. On the other hand, the $m_1$ distribution does not provide a more stringent limit on the lower end of $M_{\rm max}$  for $\alpha$ close to 3.0. So, the $m_1$ and $z$ distributions provide complimentary information for us to constrain or eventually measure the mass-spectrum parameters. 

Combining both distributions and based on the observed five events, we have 
\beqa
&& 36 \,M_{\odot} \lesssim  M_{\rm max} \lesssim 90 \, M_{\odot}\,, \qquad \mbox{for} \qquad \alpha=2.3 \,,   \\
&& 36 \,M_{\odot} \lesssim  M_{\rm max} \lesssim 60 \, M_{\odot}\,, \qquad \mbox{for} \qquad \alpha=1.6 \,,  \\
&& 42 \,M_{\odot} \lesssim  M_{\rm max} \lesssim \infty \,, \qquad \quad \;\;\,\mbox{for} \qquad \alpha=3.0 \,, 
\eeqa
for $M_{\rm min}=5\,M_\odot$ and $\beta=0$. Some of the lower limits  with $36 \,M_{\odot}$  come from the observed mass of the event GW150914. Although the parameter $M_{\rm min}$ is fixed in our calculations, we have checked and found that the constraints on $M^{\rm max}$ are insensitive to the actual values of $M_{\rm min}$, once it is much smaller than $M_{\rm max}$.

\section{Future sensitivity to measure the mass cap}
\label{sec:future}

\begin{figure}[htb!]
\begin{center}
\includegraphics[width=0.48\textwidth]{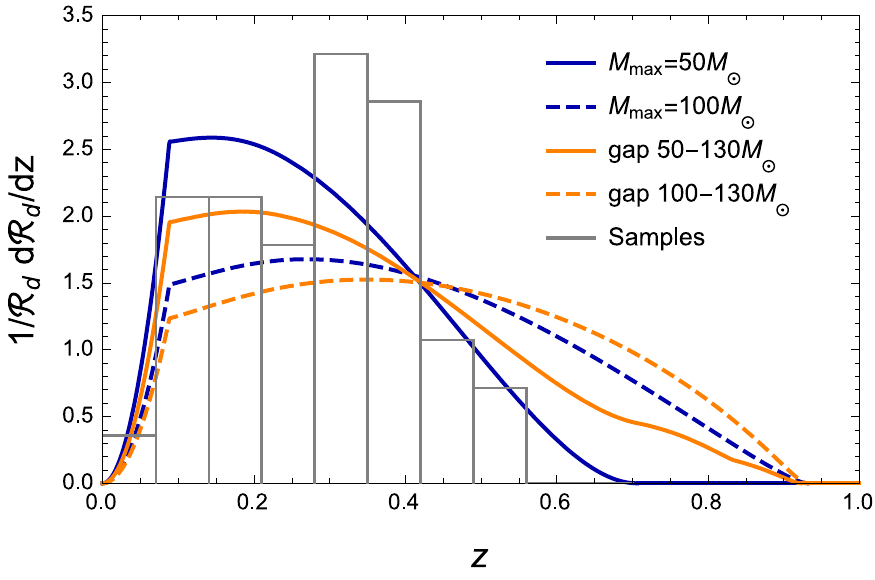}
\hspace{3mm}
\includegraphics[width=0.48\textwidth]{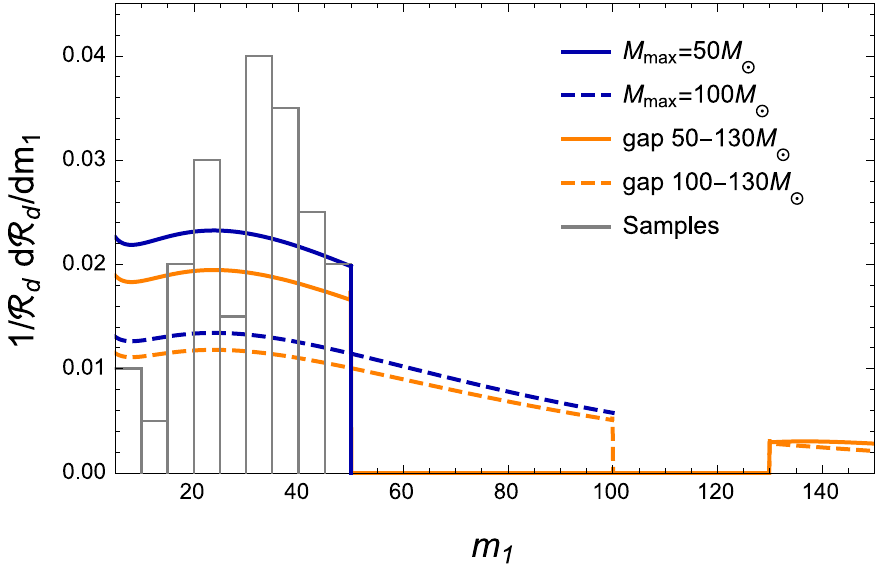}
\vspace*{-0.3cm}
\caption{Left panel: the normalized $d\mathcal{R}_d/dz$ distributions for $\alpha=2.3$ and different mass caps or gaps. The gray histogram is a sample of 40 pseudo events based on a mass-spectrum model of $\alpha=2.3$ and $M_{\rm max}=50\,M_\odot$. Here, $M_{\rm min}=5\,M_\odot$ and $\beta=0$. Right panel: the same as the left one but for $d\mathcal{R}_d/dm_1$ distributions.}
\label{fig:future}
\end{center}
\end{figure}

After obtained the constraints on the mass cap from the observed five events, we briefly estimate the future sensitivity on constraining the mass-spectrum parameters. As the aLIGO detectors will undergo further upgrades to reduce the background noise, it is interesting to see how the mass cap or gap influences the redshift distribution of the future observed events and how well the cap or gap can be measured with a better sensitivity. The future strain noise is expected to be smaller than the current minimum value by around a factor of two, over a wider frequency range of about $50\sim500$~Hz \cite{ligo:noise_document}. In this section we use the analytic expression for the strain noise \cite{Ajith:2011ec}
\beqa
S_n(f) = 10^{-48}\,(0.0152 \, x^{-4}+0.2935 \, x^{9/4}+2.7951 \, x^{3/2}-6.5080 \, x^{3/4}+17.7622)\,, 
\eeqa
which matches the advanced LIGO final design very well. Here, $x\equiv f/245.4\,\mbox{Hz}$. With this noise strain template and requiring SNR above 8, we calculate the maximum redshift $z_{\rm max}(m_1, m_2)$, which has the form of
\begin{align}
z^{\rm future}_{\rm max}(m_1, m_2) &\approx \exp\left\{-3.84 + 0.70\,\ln{\left[\frac{(m_1\,m_2)^{0.92}}{(m_1 + m_2)^{0.41}}\right] } \right.\nonumber \\
& \hspace{1cm} \left.-10^{-6}\times\left[5.50+1.84\left(\frac{m_1}{m_2}+\frac{m_2}{m_1}\right)^{0.72}\right]\,\ln^{6.1}{\left[\frac{(m_1\,m_2)^{0.92}}{(m_1 + m_2)^{0.41}}\right] } \right\} \,,
\end{align}
for $m_1$ and $m_2$ in the unit of $M_\odot$.

After integrating out $m_2$ and $m_1$, we show the redshift distribution of model-predicted events, $d\mc{R}_d/dz$, in the left panel of Figure~\ref{fig:future}. Other than the power-law mass spectrum with a cap, $M_{\rm max}=50(100)\,M_\odot$, we also show the distributions for the power-law mass spectrum with a gap, $50-130\,M_\odot$ and $100-130\,M_\odot$. One can see that the change of distributions caused by the higher end of the gap is not as significant as the influence of increasing $M^{\rm low}_{\rm gap}$ or $M_{\rm max}$. This is simply due to power-law behavior of the mass spectrum: there are fewer black holes with a heavier mass. So, if $M^{\rm high}_{\rm gap} \gtrsim 100\,M_\odot$, it will require even more events to measure the higher end of the gap. However, for the lower end of the gap, a few dozen of events may be enough to constrain $M^{\rm low}_{\rm gap}$ or $M_{\rm max}$ to be within a 10 solar mass accuracy. Therefore, we choose a mass spectrum model with $\alpha=2.3$ and $M_{\rm max}=50\,M_\odot$, simulate 40 pseudo events and show the histogram distribution also in the left panel of Figure~\ref{fig:future}. 

\begin{figure}[htb!]
\begin{center}
\includegraphics[width=0.6\textwidth]{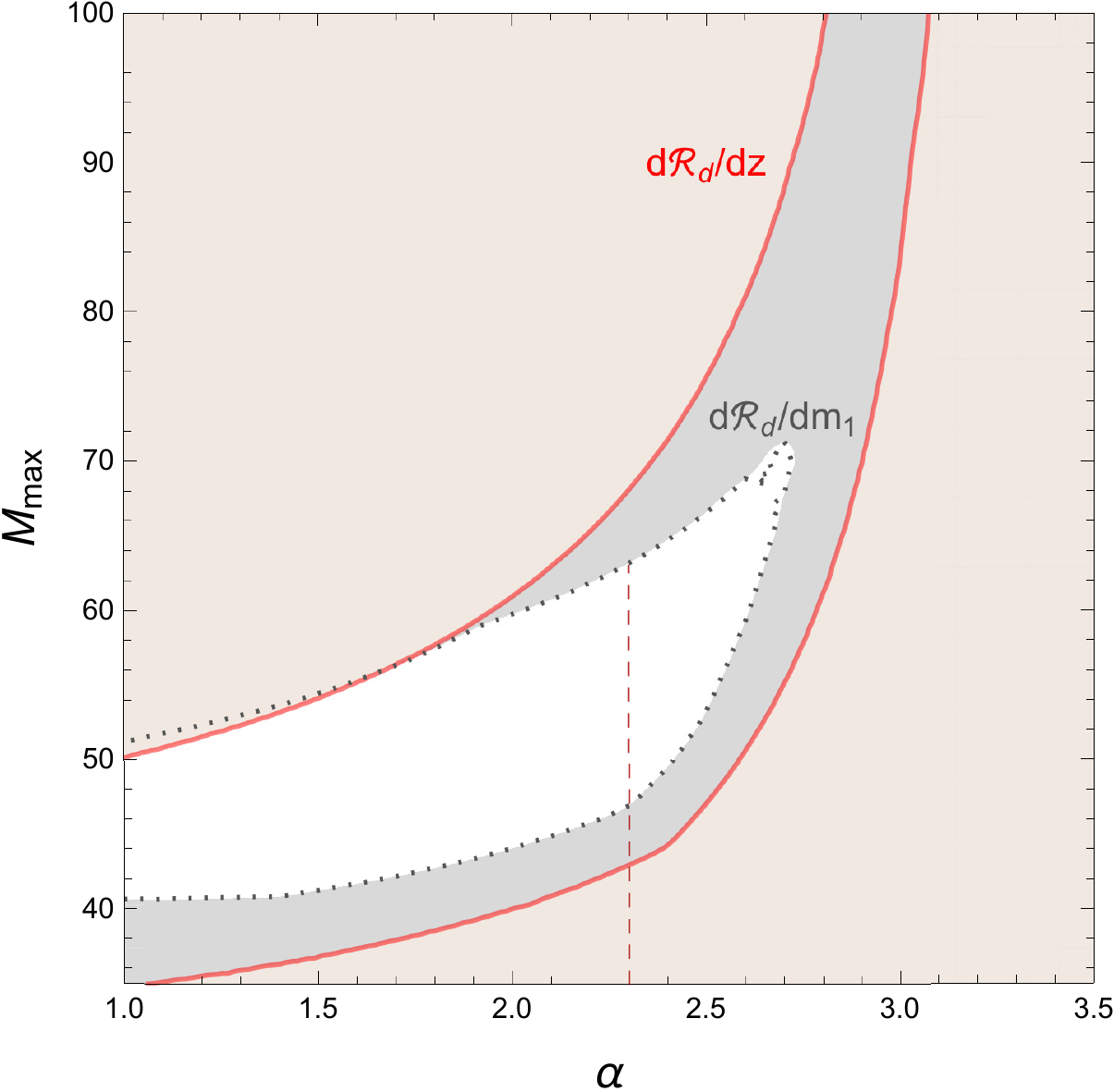}
\vspace*{-0.3cm}
\caption{Same as Figure~\ref{fig:p-value}, but based on the future sensitivity of aLIGO and 40 pseudo events from the model with $\alpha=2.3$ and $M_{\rm max}=50\,M_\odot$. }
\label{fig:40}
\end{center}
\end{figure}

Comparing the left panel of Figure~\ref{fig:future} to Figure~\ref{fig:z-distribution}, one can see that we should anticipate more observed events with a higher range of redshift from $\sim 0.2$ to 0.4. The highest value of possible redshift is also extended to be around 0.7 from 0.23 with the current sensitivity, for $\alpha=2.3$ and $M_{\rm max}=50\,M_\odot$. For a higher value of $M_{\rm max}$, the extension of the $z$-distribution to higher values is more dramatically than the current situation. So, we anticipate that the $d\mathcal{R}_d/dz$ distribution will become more useful to constrain $M_{\rm max}$. In the right panel of Figure~\ref{fig:future}, we show the normalized $d\mathcal{R}_d/dm_1$ distributions for different mass-spectrum models as well as for the sample of 40 pseudo events . One can see that it is easy to distinguish the two models with $M_{\rm max}=50\,M_\odot$ and $M_{\rm max}=100\,M_\odot$, and more difficult to measure the upper end of the mass gap.

Using the KS test, we show the 95\% confidence level constraints on the mass-spectrum parameters, $\alpha$ and $M_{\rm max}$, in Figure~\ref{fig:40}, based on the 40 pseudo events from the model with $\alpha=2.3$ and $M_{\rm max}=50\,M_\odot$. One can see that both $d\mathcal{R}_d/dz$ and $d\mathcal{R}_d/dm_1$ distributions provide comparable constraints on the model parameters and the $d\mathcal{R}_d/dm_1$ distribution is slightly more constraining. For the central power-law index $\alpha=2.3$, one can constrain the model-parameter, $M_{\rm max}$, to be within a window of around $10\,M_{\odot}$. As we discussed below Figure~\ref{fig:future}, the mass spectrum parameter, $M^{\rm high}_{\rm gap}$, does not influence distributions as significantly as the $M_{\rm max}$ parameter of the mass-cap models. So, more events are required to constrain or measure $M^{\rm high}_{\rm gap}$, so we don't report the results of pseudo experiments based on those mass spectra.

\section{Analysis with ten events}
\label{sec:ten-events}

On November 30, 2018, the LIGO and Virgo collaboration announced four new BBH merger events in the second observing run, with the previously less-confident event LVT151012 also formally recognized as a BBH merger event GW151012~\cite{LIGOScientific:2018mvr}. We, therefore, update our analysis using the totally around 10 events. 

\begin{figure}[htb!]
\begin{center}
\includegraphics[width=0.6\textwidth]{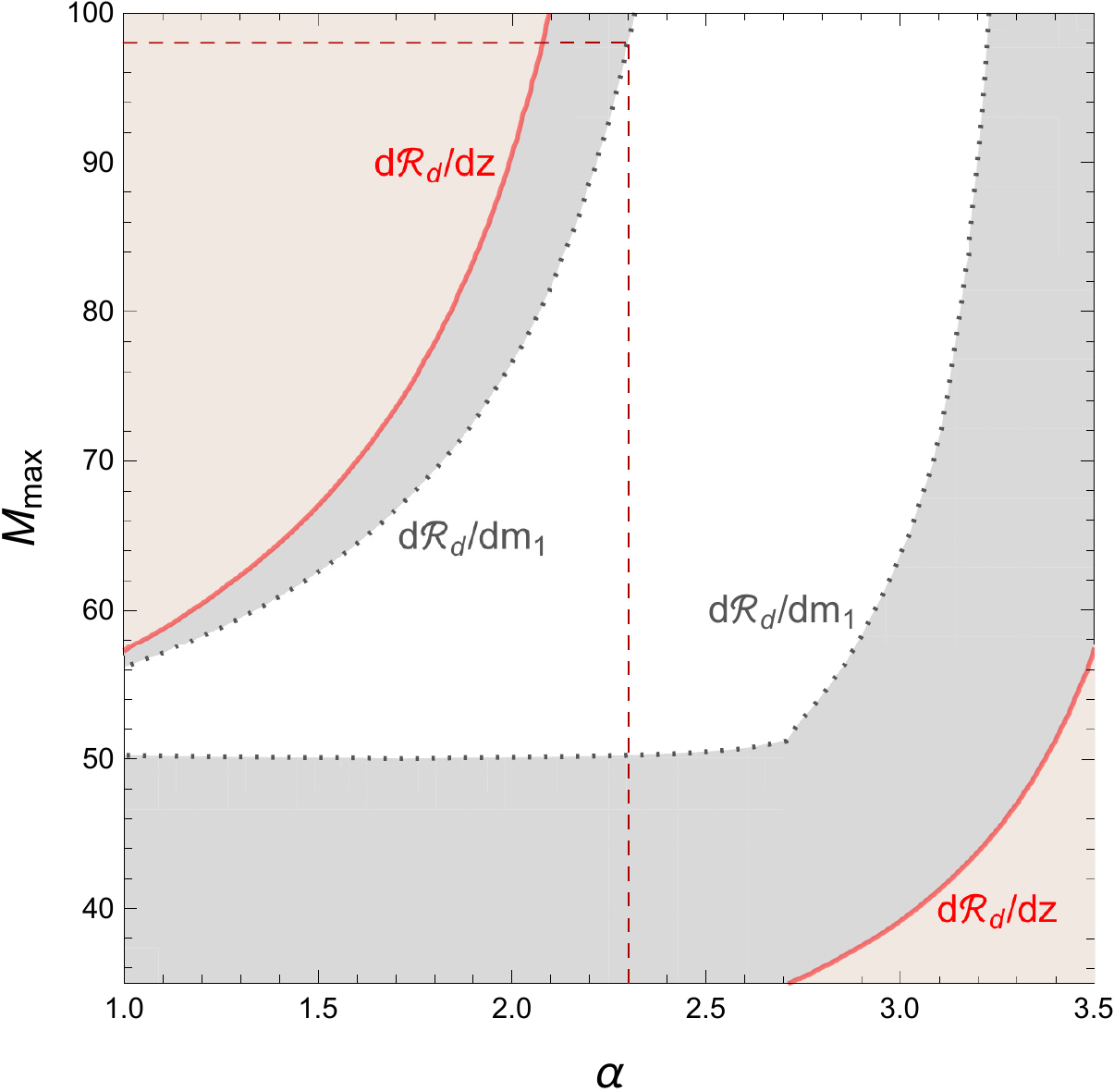}
\vspace*{-0.3cm}
\caption{The 95\% confidence level from the $d\mathcal{R}_d/dz$ distribution (light brown region) and $d\mathcal{R}_d/dm_1$ distribution (light gray region) of the ten observed BBH merger events summarized in \cite{LIGOScientific:2018mvr}. The horizontal line in the $d\mathcal{R}_d/dm_1$ exclusion region is due to the 50.6~$M_\odot$ black hole observed in the event GW170729, which suggests $M_{\rm max}\gtrsim50.6$~$M_\odot$ by existence.}
\label{fig:10}
\end{center}
\end{figure}

Instead of using the SNR of a single detector to impose a detectability threshold, for this section we switch to the SNR of the detector network such that all the events in Table~\ref{tab:data} are observable. The new threshold used in the analysis of this section is
\begin{align}
\left(\frac{S}{N}\right)_{\rm net}=\sqrt{\left(\frac{S}{N}\right)^2_{\rm H}+\left(\frac{S}{N}\right)^2_{\rm L}  } > 6  ~,
\end{align}
where the subscripts ``H" and ``L" stand for the SNR calculated at the Hanford and Livingston detector, respectively. As the strain noise of the Virgo detector is much higher than the other two~\cite{Abbott:2017oio,LIGOScientific:2018mvr}, we do not include it into our calculation of SNR. The strain noises of the two LIGO detectors, on the other hand, change over the two observing runs. To calculate the SNR of the two individual detectors, we perform a fit to all ten strain noise profiles and obtain an averaged background. The parametrized strain noises are given by
\begin{align}
\sqrt{S_{n, {\rm H}}}=7.86\times10^{-26}\,f^{-1.68}\,e^{\frac{31.74}{\ln f}+0.26\ln^2 f} \,,\quad
\sqrt{S_{n, {\rm L}}}=2.70\times10^{-37}\,f^{4.39}\,e^{\frac{66.62}{\ln f}-0.17\ln^2 f}  \,,
\end{align}
where $f$ is in Hz. In Fig.~\ref{fig:10} we follow the same procedure as in Section~\ref{sec:five-events} and derive the 95\% confidence level constraints on $M_{\rm max}$ and $\alpha$ using the KS test. Since there is a 50.6~$M_\odot$ black hole observed in the BBH merger event GW170729, we no long consider the parameter space with $M_{\rm max} < 50.6~M_\odot$. Compared with Fig.~\ref{fig:p-value}, the constraint from the $d\mathcal{R}_d/dz$ distribution improves as the observed events accumulates. However, it is now less constraining than $d\mathcal{R}_d/dm_1$, which is mainly due to the existence of the observed heavy black holes. The constraint from the $d\mathcal{R}_d/dm_1$ distribution is also pushed toward the direction of a larger $M_{\rm max}$, {\it e.g.}, for $\alpha=2.3$ we now have $M_{\rm max}<98\,M_\odot$ instead of $90\,M_\odot$. 

It is worth to point out that the constraints obtained from the two distributions can be influenced by the statistical method adopted in the analysis. One may also perform a Bayesian posterior inference, instead of using the relatively straightforward comparison of the data samples with the theoretical distribution, as in the KS test. As a comparison, we have also performed an analysis based on Bayesian posterior inference in Appendix~\ref{sec:bayesian}. In fact, the result from our simple posterior inference shows that the redshift distribution can still be complimentary to the mass distribution for constraining some model parameter space, which is aligned with the main point of the paper. 
\section{Discussion and conclusions}
\label{sec:conclusion}
In our study, we used a semi-analytical approximation to the general relativity templates of binary black hole signals and a fit to the smooth component of the aLIGO detector noise. We employ parameterizations of the frequency dependence of the signal and of the noise [see Eqs.~\eqref{eq:noise-fit}\eqref{eq:inspiral-merge}]. Although our approximate approach is less precise than the use of numerical templates provided by aLIGO, it provides a framework that could be more transparent to the underlying physics, and thus it may also be very useful for future study of  possible new physics effects. We obtained 95\% confidence level constraints on the black hole mass distribution parameters using the KS test on the $d\mathcal{R}_d/dz$ and $d\mathcal{R}_d/dm_1$ distributions separately. Our study showed that the $d\mathcal{R}_d/dz$ distribution provided constraints that are complementary to those obtained from the $d\mathcal{R}_d/dm_1$ distribution. Eventually, one should combine both distributions or even construct a 2-dimensional KS test to obtain the final constraints on $\alpha$ and $M_{\rm max}$.

For the BBH merging events, we have explored black hole mass distribution models with factorable redshift and mass dependences, or $f(m_1, m_2, z)=p(m_1, m_2) R(z)$, in \eqref{eq:triple-diff}, similarly to other studies in the literature~\cite{Kovetz:2016kpi,Fishbach:2017zga,Talbot:2018cva}. The realities could be more complicated than this. For instance, the numerical studies in Ref.~\cite{Belczynski:2009xy} have shown that the mass-cap parameter, $M_{\rm max}$, in $p(m_1, m_2)$ depends on the metallicity, and hence redshift, of the progenitors of the black holes. With further events from aLIGO and VIRGO, it should be possible to test a more complex interplay of such physics.

In conclusion, we showed that the observed redshift distribution of the known BBH merger events constrain the parameterization of the black hole mass function. From the observed five events, we constrained the maximum value of the black hole masses versus the power-law index of the black hole mass spectrum.  For example, we find $M_{\rm max} < 90\,M_\odot$ for a negative power-law index of $\alpha=2.3$ for the heavier black hole and $M_{\rm max} < 60\,M_\odot$ for a flatter $\alpha=1.6$. After the detector noise is reduced in the aLIGO detector upgrade, it is anticipated that a few dozen BBH merger events will be observed leading to tighter constraints on the mass distribution. Moreover, the first observations of a neutron star-black hole merger will provide supplementary information on the black hole mass function and contribute to our understanding of the metallicity of the environment in which the compact objects formed.

\vspace{1.5cm}
{\it  Acknowledgments:}  {\small 
The work  is supported by the U. S. Department of Energy under the contract DE-SC0017647. 
}
\begin{appendix}
\section{Bayesian posterior analysis}
\label{sec:bayesian}
\begin{figure}[t!]
\begin{center}
\includegraphics[width=0.6\textwidth]{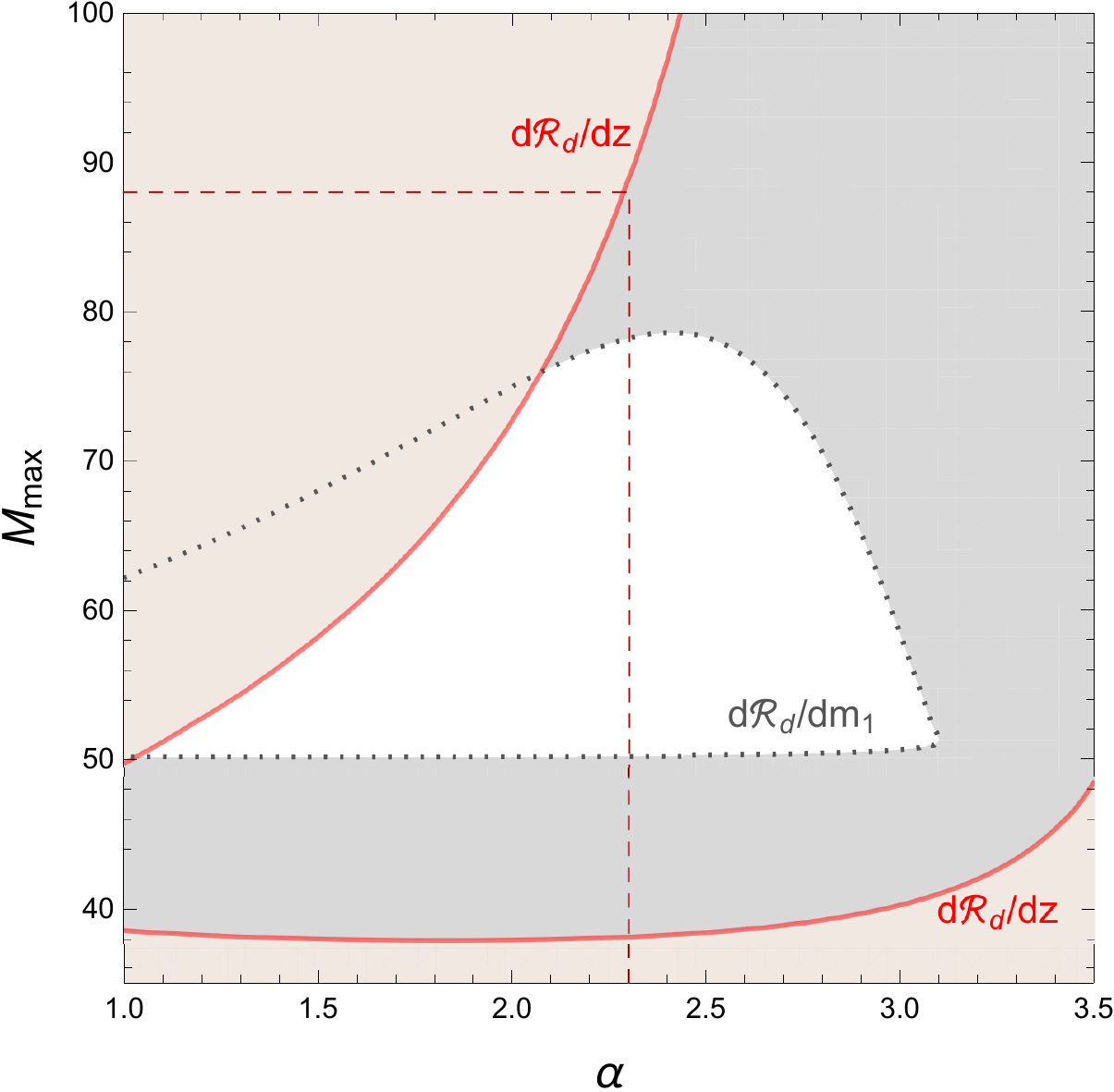}
\caption{The 95\% likelihood parameter space inferred from the redshift and mass distributions, based on the 10 BBH events in \cite{LIGOScientific:2018mvr} with Bayesian posterior inference. The threshold value of the network SNR is chosen to be 6.}
\label{fig:contour_posterior}
\end{center}
\end{figure}
In this appendix, we perform Bayesian inference on the parameter $\alpha$ and $M_{\rm max}$ and compare the result obtained from the KS test. Given the measurements, the posterior probability distribution function (PDF) of the parameters is calculated by
\begin{align}
\label{eq:post1}
p(\vec{\theta}\,|\,\vec{z},\,\vec{S}_n) = \frac{p(\vec{z}\,|\,\vec{\theta},\,\vec{S}_n)\, \pi(\vec{\theta})}{\int d\vec{\theta}\,p(\vec{z}\,|\,\vec{\theta},\,\vec{S}_n)\, \pi(\vec{\theta})}  ~,
\end{align}
where $\vec{\theta}=(\alpha,\,M_{\rm max})$ is the parameter vector for inference; $\vec{z}$ and $\vec{S}_n$ are the measured redshift and strain noise of each measurement; $\pi(\vec{\theta})$ is the prior distribution of the parameters. In the following analysis, we take a flat prior over the parameter space $\alpha\in[1.0,\,3.5],\,M_{\rm max}\in [35\,M_\odot,\,100\,M_\odot]$. Thus, we have
\begin{align}
\label{eq:post2}
p(\vec{\theta}\,|\,\vec{z},\,\vec{S}_n) \propto p(\vec{z}\,|\,\vec{\theta},\,\vec{S}_n)=\prod_i p_i (z_i\, | \,\vec{\theta},\, S_{n,\,i}) ~,
\end{align}
where in the second equation we treat the measurements to be independent of each other, and the index $i$ runs over the measurements. The probability $p_i (z_i\, | \,\vec{\theta},\, S_{n,\,i})$, similar to the expression in Eq.~\eqref{eq:triple-diff}, is given by
\begin{align}
\label{eq:post3}
p_i (z_i\, | \,\vec{\theta},\, S_{n,\,i}) = \frac{1}{R_i}\int dm_1\,dm_2\, p(m_1, m_2)\,R(z)\, \frac{d V_c}{dz} \frac{dt_s}{dt} \, \Theta[z_{{\rm max}, i}(m_1, m_2) - z] ~,
\end{align}
with $R_i$ as the normalization factor. The posterior PDF of the primary mass distribution, $p_i (m_{1,i}\, | \,\vec{\theta},\, S_{n,\,i})$, can be obtained in a similar way, after marginalizing over $z$ instead of $m_1$. Combining the equations above, the PDF can be calculated over the parameter space of interest. In Fig.~\ref{fig:contour_posterior} and using all 10 confirmed events, we show the 95\% likelihood regions inferred from the posterior PDF of both measured redshift and primary mass distributions. 

Comparing the regions obtained from the KS test in Fig.~\ref{fig:10} and from the Bayesian posterior inference in Fig.~\ref{fig:contour_posterior}, one can see that the allowed parameter space inferred from the Bayesian approach is smaller. One may wonder why the results based on the $d\mathcal{R}/dm_1$ distributions from the two different statistical methods are so different. The simple explanation is that the Bayesian inference disfavors lager values of $M_{\rm max}$ because of the usage of a normalized probability distribution for individual events [see Eq.~\eqref{eq:post3}]. A larger $M_{\rm max}$ will decrease each $p_i (m_{1,i}\, | \,\vec{\theta},\, S_{n,\,i})$ compared to a smaller one. On the other hand and from Fig.~\ref{fig:m1-distribution}, a larger value of $M_{\rm max}$ is preferred for the KS test when $\alpha$ is large, because  the cumulative distribution function is ``stretched" to fit better to the stair-like distribution of the samples.

\end{appendix}
\bibliography{BBH}
\bibliographystyle{JHEP}

 \end{document}